\documentclass[a4paper,11pt]{article}
\usepackage{epsfig}

\begin{document}

\begin{center}

{\Huge Critical Temperature for the Nuclear Liquid-Gas Phase Transition}\\

\large{V.A.~Karnaukhov$^1\footnote{email: karna@nusun.jinr.ru}$},
H.~Oeschler$^2$,
S.P.~Avdeyev$^1$,  E.V.~Duginova$^1$,
V.K.~Rodionov$^1$, A.~Budzanowski$^3$,  W.~Karcz$^3$,
O.V.~Bochkarev$^4$, E.A.~Kuzmin$^4$, L.V.~Chulkov$^4$,
E.~Norbeck$^5$, A.S.~Botvina$^6$ \\
{\small\it $^1$ Joint Institute for Nuclear Research, 141980 Dubna, Russia} \\
{\small\it $^2$ Institut f\"ur Kernphysik, University of Technology,
64289 Darmstadt,  Germany}\\
{\small\it $^3$ H.~Niewodniczanski Institute of Nuclear Physics, 31-342,Cracow,
Poland }\\
{\small\it $^4$ Kurchatov Institute, 123182, Moscow, Russia}\\
{\small\it $^5$ University of Iowa, Iowa City, IA 52242 USA  }\\
{\small\it $^6$ GSI, Postfach 110552, D-64220 Darmstadt, Germany}
\end{center}

\section{}
The charge distribution of the intermediate mass fragments produced in p (8.1
GeV) + Au collisions is analyzed in the framework of the statistical
multifragmentation model with the critical temperature for the nuclear
liquid-gas phase transition $T_c$ as a free parameter. It is found that $T_c=20\pm3$~MeV (90\% CL).

{\it PACS} : 24.10.Lx, 25.70.Pq

\newpage
\section{}
The extensive study of nuclear multiframentation for the last two decades has
been strongly stimulated by the idea that this process is related to a
liquid-gas phase transition. One of the first nuclear
models, suggested by N.~Bohr, K.~Weizs\"acker and Ya.I.~Frenkel 65 years ago,
is the liquid-drop model, which is still alive. A liquid-gas phase transition
in nuclear matter was predicted much later [1-3] on the basis of the
similarity between van der Waals and nucleon-nucleon interactions~\cite{2}.
The equations of state for the two cases are similar. A point of particular
interest is the existence of a spinodal region at reduced densities
characterized by a phase instability. One can imagine that a hot nucleus
expands due to thermal pressure and enters into this unstable region.
Due to density fluctuations, a homogeneous system converts into a mixed 
phase, consisting of droplets (IMF's, $2<{\rm Z}\le20$) surrounded by
nuclear gas. In fact, the final state of this transition is a {\it nuclear
fog}~\cite{3}. The neutrons fly away with energies corresponding to the
system temperature ($6-7$~MeV), while the charged particles are additionally
accelerated by the Coulomb field of the system. The disintegration time is
determined by the time scale of the density fluctuations and is very short.
Indeed, it was measured in number of papers that the IMF's emission time is less than 100 fm/{\it c}.
This is the scenario of nuclear multifragmentation as a spinodal
decomposition, considered in a number of papers
(see, for example, [4-12], and review papers [13-14]). The spinodal
decomposition is, in fact, the {\it liquid-fog} phase transition in a nuclear 
system.

An important parameter of this scenario is the critical
temperature for the nuclear liquid-gas phase transition $T_c$ at which the isotherm in
the phase diagram has an inflection point. The surface tension vanishes at
$T_c$, and only the gas phase is possible above this temperature. There are
many calculations of $T_c$ for finite nuclei. In
Ref.~\cite{1,2,15,16}, for example, it is done by using a Skyrme interaction and the thermal Hartree-Fock theory. The values of $T_c$ were
found to be $10-20$~MeV, depending on the Skyrme
interaction parameters and the details of the model. Experimental estimations of the critical temperature for the finite nuclei have been done in several papers.

The main source of experimental information for $T_c$ is the fragment
yield, but the procedures to extract $T_c$ are heavily debated. In some statistical models of nuclear multifragmentation the shape of
the IMF charge distribution {\it Y}(Z) is sensitive to the ratio
$T/T_c$. The charge distribution is well described by the power law $Y({\rm
Z})\sim{\rm Z}^{-\tau}$ for a wide range of colliding systems~\cite{17}.
In earlier studies on multifragmentation~\cite{3,18} the power-law behavior
of the IMF yield was interpreted as an indication of the proximity of the
excited system to the critical point for the liquid-gas phase transition. This was stimulated by the application of Fisher's classical
droplet model~\cite{19}, which predicted a pure power-law droplet-size
distribution with the minimal value of $\tau=2-3$ at the critical point.

In Ref.~\cite{18} Hirsch et al. estimate $T_c$ to be $\sim 5$~MeV simply from
the fact that the mass distribution is well described by a power law for
IMF's produced in the collision of p ($80-350$~GeV) with Kr and Xe targets. In fact, the fragment mass distribution is not exactly described by the power
law, therefore it was suggested the use of the term
$\tau_{app}$, an apparent exponent, to stress that the exact power-law
description takes place only at the critical temperature. In paper~\cite{20}
the experimental data were gathered for different colliding systems to get
the temperature dependence of $\tau_{app}$. As a temperature, the inverse
slope of the fragment energy spectra was taken in the range of the
high-energy tail. The minimal value of $\tau_{app}$ was obtained at 
{$T=11-12$ MeV, which was claimed as $T_c$. The later data smeared out this minimum. Moreover, it became clear that the ``slope" temperature does not coincide with the thermodynamical one which is several times
smaller.

A sophisticated use of Fisher's droplet model for the estimation of
$T_c$ has been recently made by Elliott, Moretto et al.~\cite{21,22}. The
model was modified by including the Coulomb energy release when a particle
moves from the liquid to the vapor. The data from the
Indiana Silicon Sphere Collaboration for $\pi$ (8 GeV/c) + Au collisions were
analyzed~\cite{21}. The extracted critical
temperature is $T_c=6.7\pm0.2$ MeV. In a recent paper~\cite{22} the same
analysis technique is applied to the data for the multifragmentation in
collisions of Au, La, Kr (at 1.0 GeV per nucleon) with a carbon target. The extracted values of $T_c$ are $7.6\pm0.2$,
$7.8\pm0.2$ and $8.1\pm0.2$~MeV respectively.

There is only one paper in which $T_c$ is estimated by using data other than the fragmentation ones. In Ref.~\cite{23} it is done by the
analysis of the temperature dependence of the fission probability for
$^4$He + $^{184}$W collisions~\cite{24}. It was concluded that $T_c>10$ MeV in
contrast to the result of Ref.~\cite{21,22}.

It should be noted that in some papers the term ``critical temperature" is
 used in another meaning than given above. In Ref.~\cite{25}
multifragmentation in Au + Au collisions at 35 A MeV was analyzed with the
so-called Campi plots to prove that the phase transition takes  place in the
spinodal region. The characteristic temperature for that process was denoted
as $T_{crit}$ and found to be equal to $6.0\pm0.4$~MeV. In the recent
paper~\cite{26} the bond percolation model is used to interpret 10.2 GeV/{\it
c} p + Au multifragmentation data. The critical value
of the percolation parameter $p_c=0.65$ was found from the analysis of the
IMF charge distribution. The corresponding ``critical temperature" of 
$8.3\pm0.2$~MeV is estimated by using the model relation between the
percolation control parameter ``{\it p}" and the excitation energy.
The more appropriate term for this particular temperature is ``break-up" or ``crack" temperature, as suggested in Ref.~\cite{27}. This temperature corresponds to onset of the fragmentation of the nucleus entering the phase coexistence region. The low-multiplicity channels dominate during the onset of multifragmentation characterized by a U-shaped fragment mass distribution. As shown by means of the statistical multifragmentation model (SMM)~\cite{27,28}, the average hot fragment multiplicity is M=3-5 at an excitation energy around 4~MeV/nucleon, and the probability of the compound nucleus channel is still considerable. Exactly at these relatively low excitation energies the experimenters observed the critical phenomena (see, for example~\cite{25,29,30}).

Having in mind the shortcomings of Fisher's model~\cite{31,32}, we have made
an attempt to estimate the critical temperature in the framework of SMM. It
describes well the properties of the thermal fragmentation of target
spectators produced in collisions by light relativistic ions. As an
example, Fig.~1 (top) shows the fragment charge distributions measured by
the FASA Collaboration for collisions of p (8.1~GeV), $^4$He (4 and
14.6~GeV) and $^{12}$C (22.4~GeV) with Au targets~\cite{12} along with the 
calculated charge distributions. The mechanism for the reactions of light 
relativistic projectiles is usually divided into two
stages. The first is a fast energy-depositing stage, during  which very
energetic light particles are emitted and a nuclear remnant is excited. We
use the intranuclear cascade model (INC)~\cite{33} for describing the first
stage. The second stage is described by SMM, which considers the multibody
decay of a hot and expanded nucleus. But such a two-stage approach fails to
explain the observed IMF multiplicities. An expansion stage (Exp) is inserted
between the two parts of the calculation. The excitation energies and the
residual masses are then fine tuned on an event-by-event basis~\cite{12} to get agreement with the
measured IMF multiplicities.
The lines in Fig.~1 (top) give the charge distributions calculated in the
framework of this combined model assuming $T_c=18$~MeV. The agreement
between the data and the model prediction is very good. The bottom panel of Fig.~1 shows the power-law fit of the distributions with
the $\tau$ parameter given in the insert as a function of the beam energy.
The corresponding thermal excitation energy range is $3-6$~MeV/nucleon. The
power law parameter exhibits the so-called ``critical behavior" showing a
minimum at an excitation energy corresponding to a temperature three times
lower than the assumed $T_c$. A conventional explanation of the occurrence
of a minimum is given in
Ref.~\cite{12,17}.

The charge yield depends on the contribution of the surface free
energy of the fragments ($a_s(T)A^{2/3}$) to the entropy 
of a given final state of the partition. The 
following expression is used in the SMM for $a_s(T)$:

\begin{equation}
a_s(T) = a_s(0) \Biggl(\frac{T_c^2-T^2}{T_c^2+T^2}\Biggr)^{5/4}\quad.
\end{equation}

This equation was obtained in Ref.~\cite{34} devoted to the theoretical study
of thermodynamical properties of a plane interface between two phases of
nuclear matter (liquid and gas) in equilibrium. The corresponding calculations were performed with the Skyrme interaction. The phase diagram generated by the SMM model (using eq. (1)) is discussed in detail in~\cite{35}. This parametrization is
successfully used by the SMM for describing the multifragment decay of 
hot finite nuclei. In particular the SMM describes the experimental critical behaviour of fragments and scaling in multifragmentation~\cite{25,29,30} with the standard $T_c=18$~MeV. This scaling was taken as a starting point of the analyses~\cite{21,22} also.

The present calculations are performed for p (8.1 GeV) + Au collisions with
$T_c$ as a free parameter. For all values of $T_c$ the calculations with the
INC+Exp+SMM model have been properly adjusted~\cite{12} to get the mean IMF
multiplicity close to the measured one. Figure~2 (left) shows the comparison of the measured fragment charge
distribution with the model predictions for $T_c=7$, 11 and 18~MeV. The
statistical errors of the measurements do not exceed the size of the 
dots. The calculations are close to the data for $T_c=18$~MeV. The estimated mean
temperature of the fragmenting system is around 6~MeV, the mean charge and
mass numbers are 67 and 158 respectively. The theoretical curves deviate from
the data with decreasing $T_c$. The right panel gives the results of the power-law fits for the data and
model calculations (in the range Z=$3-11$). 

The final results are
shown in Fig.~3. The measured power-law exponent is given as a band with a 
width determined by the statistical error. The size of the symbols for the
calculated values of $\tau_{app}$ is of the order of the error bar. The model predicted values of the power-law exponent are significantly smaller than the measured one for the range of $T_c<13$~MeV. From the
best fit of the data and calculations one concludes that
$T_c=20\pm3$~MeV (90\% CL).

Figure 3 shows also the results of the calculations with $a_s(T)$ linearly
dependent on $T/T_c$~\cite{21,22}:

\begin{equation}
a_s(T) = a_s(0) \Biggl(1-\frac{T}{T_c}\Biggr).
\end{equation}
 
The calculated values of $\tau_{app}$ in this case are remarkably lower than the 
measured one for any value of the critical temperature used (up to $T_c=24$~MeV).

To conclude, the IMF charge distribution for p + Au collisions at 8.1~GeV has
been analyzed within the Statistical Multifragmentation Model with $T_c$ (at
which surface tension vanishes) as a free parameter. The value
$T_c=20\pm3$~MeV (90\% CL) obtained from the best fit to the data is
considered as an effective value of the critical temperature averaged over
all the fragments produced in the collision. This value is significantly
larger than those found in Ref.~\cite{21,22} by the analysis of the
multifragmentation data in terms of Fisher's droplet formalism. A surprisingly large range of $T_c$ values in different publications indicates on the severe model dependence of the results.
Although our value for $T_c$ is model dependent, as is any other estimate of the critical temperature, the analysis presented here provides strong support for a value of $T_c>15$~MeV. Another conclusion which can be drawn from this work is that the properties of individual hot fragments, in particular, their surface energies, can be obtained from the experimental data, and they are extremely important for identification of the phase transition. This puts additional constrains on models used for description of the phase transitions in nuclear systems.

The authors are thankful to A. Hrynkiewicz, A.N. Sissakian,
S.T. Belyaev, A.I. Malakhov, N.A. Russakovich for support and to
A.V. Ignatyuk, I.N. Mishustin, V.D. Toneev, W. Reisdorf, W. Trautmann and ALADIN Workshop-2002 for illumintating
discussions. The research was supported in part by Grant No 00-02-16608 from
the Russian Foundation for Basic Research, by the Grant of the Polish
Plenipotentiary in JINR, by Grant NATO PST.CLG.976861, by Grant No1P03 12615
from the Polish State Committee for Scientific Research, by Contract No
06DA453 with Bundesministerium f\"ur Forschung und Technologie and by the US
National Science Foundation.

\newpage
\begin{center}
\begin{figure}
\epsfig{file=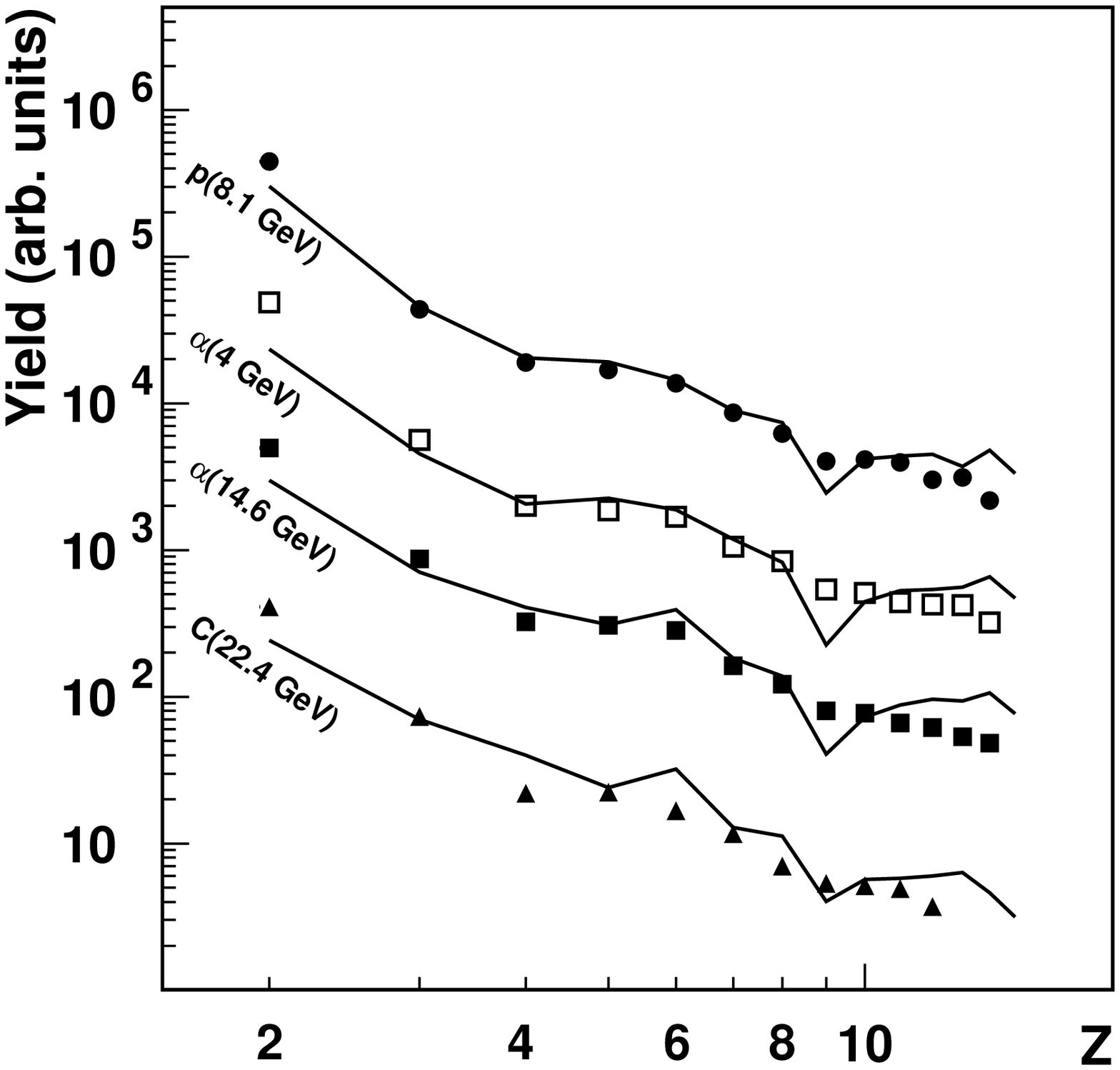,width=7.5cm} \\
\epsfig{file=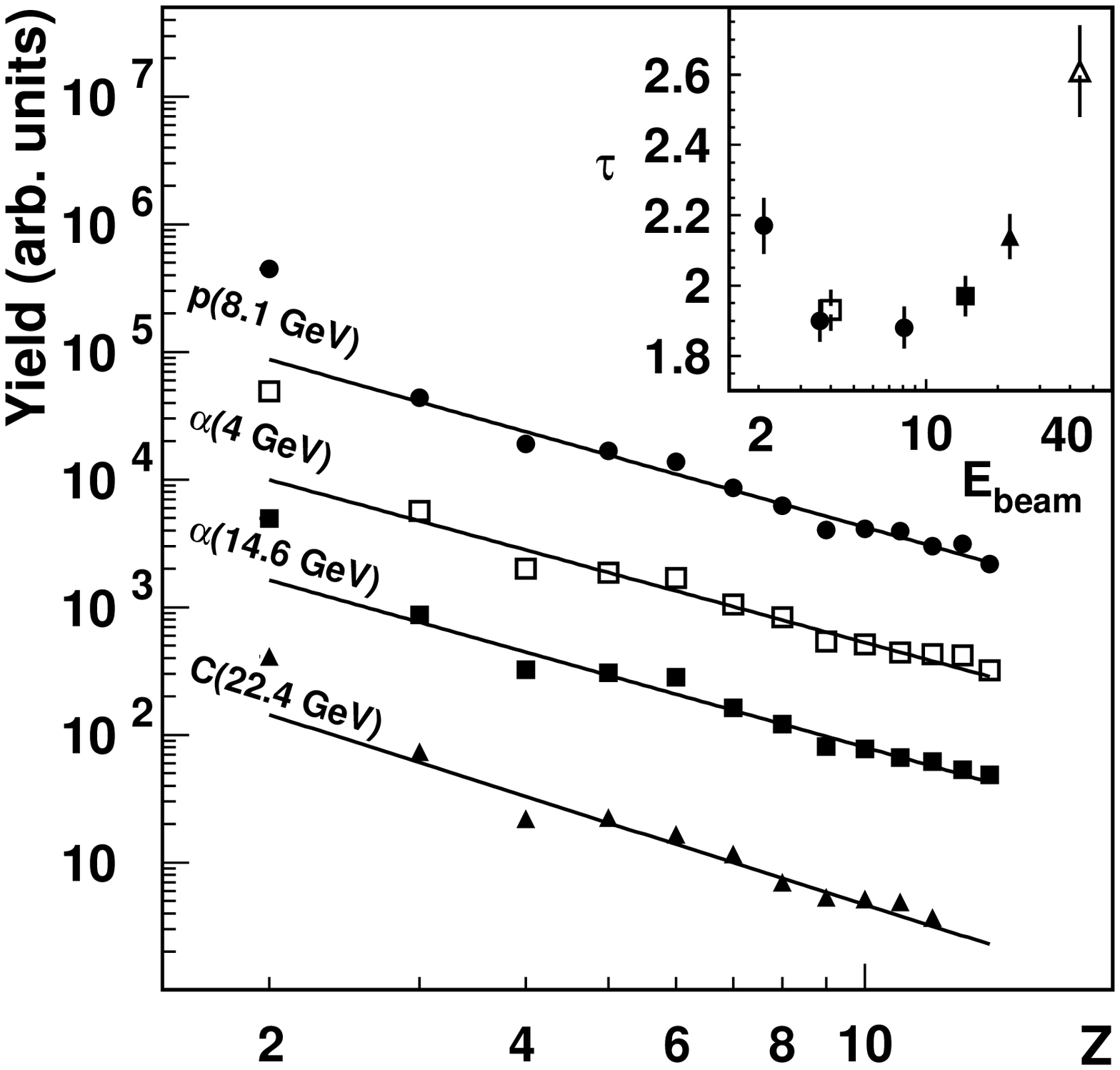,width=7.5cm}
\vspace*{-0.5cm}
\end{figure}
Figure 1. Karnaukhov, PRC.
\clearpage
\begin{figure}
\epsfig{file=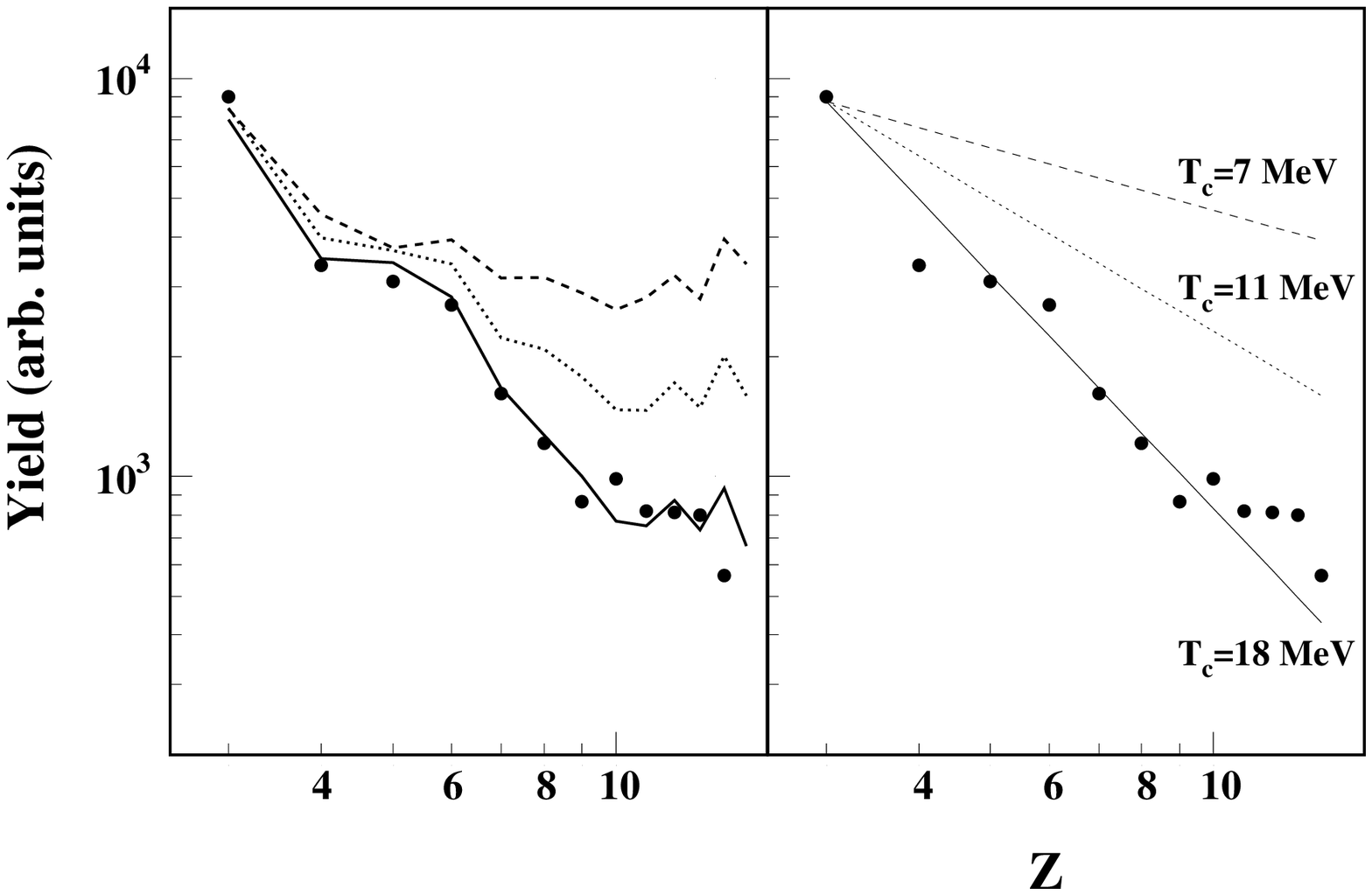,width=13cm}
\end{figure}
Figure 2. Karnaukhov, PRC.
\clearpage
\begin{figure}
\epsfig{file=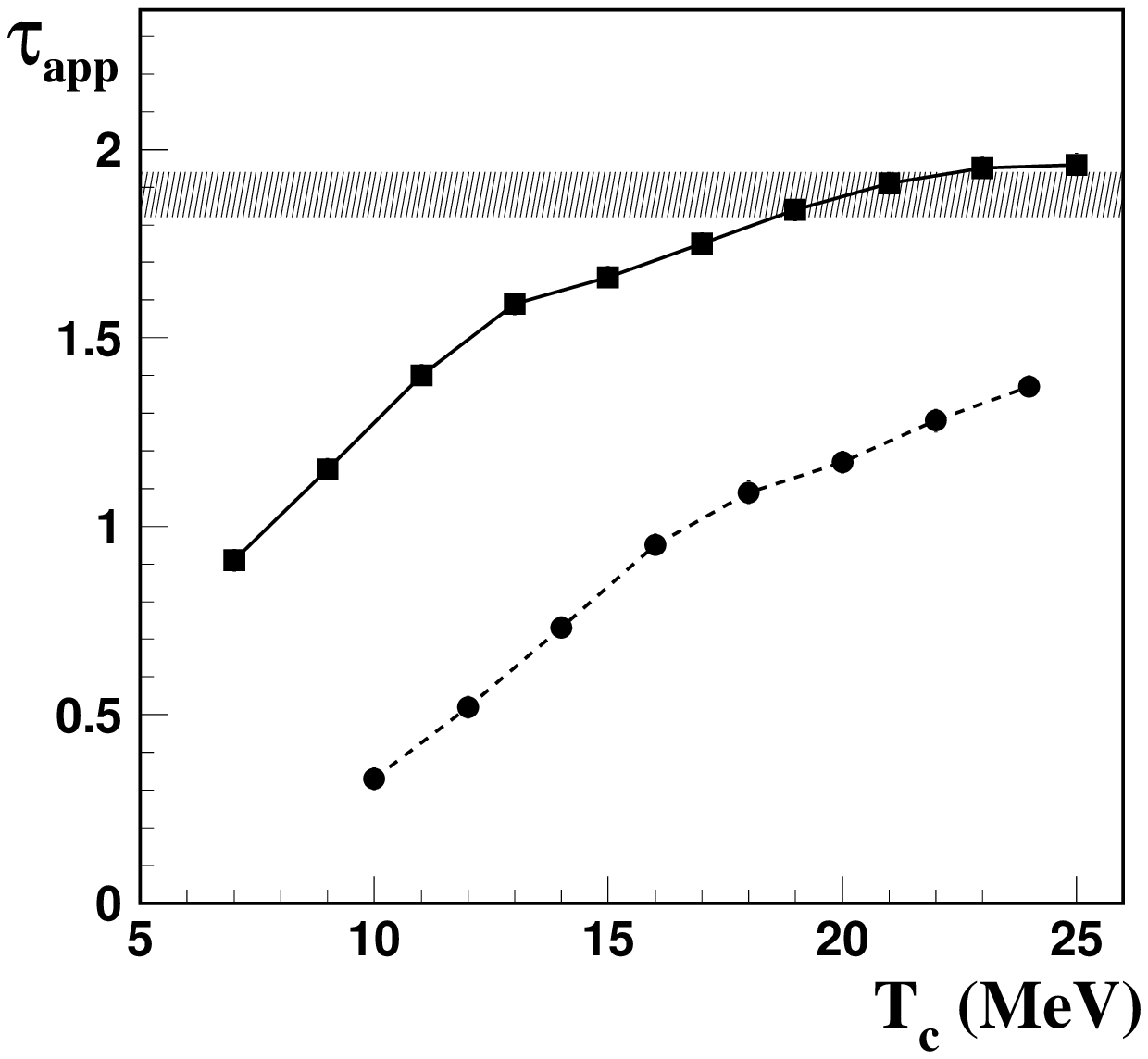,width=9cm}
\end{figure}
Figure 3. Karnaukhov, PRC.
\clearpage
Figure Captions\\
\end{center}
Fig.~1. Fragment charge distributions for p+Au at 8.1
GeV, $^4$He+Au at 4 GeV, $^4$He+Au at 14.6 GeV and $^{12}$C+Au at 22.4
GeV. The lines (top) are calculated and normalized at
Z=3. The power-law fits are shown on the bottom panel with $\tau$ parameters
given in the insert as a function of beam energy.\\

Fig.~2. Fragment charge distribution for p+Au at 8.1
GeV (dots). The lines (left side) are calculated assuming $T_c$ = 18 MeV (solid), 11 MeV (dotted) and 7 MeV (dashed lines). The
power law fits are presented on the right panel.\\

Fig.~3. The power-law exponent for p(8.1 GeV)+Au. The band corresponds to
the measured value and its error bar. The symbols are obtained by the
power-law fits of IMF charge distributions calculated assuming different
values of $T_c$ and different parameterizations of the surface tension: squares
are for eq. (1), solid circles are for eq. (2).

\begin{thebibliography}{}
\bibitem{1}  G. Sauer, H. Chandra, and U. Mosel, Nucl. Phys. {\bf A264}, 221 (1976).
\bibitem{2}  H. Jaqaman, A.Z. Mekjian, and L. Zamick, Phys. Rev. C {\bf 27}, 2782 (1983).
\bibitem{3}  P.J. Siemens, Nature {\bf 305}, 410 (1983); Nucl. Phys. {\bf A428}, 189c (1984).
\bibitem{4}  A. Guarnera, B. Jacquot, Ph. Chomaz, and M. Colonna, XXXIII Winter
Meeting on Nucl. Phys., Bormio, 1995; Preprint GANIL P 95-05, Caen,1995.
\bibitem{5}  S.J. Lee and A.Z. Mekjian, Phys. Rev. C {\bf 56}, 2621 (1997).
\bibitem{6}  V. Baran, M. Colonna, M. Di Toro, and A.B. Larionov, Nucl. Phys.
{\bf A632}, 287 (1998).
\bibitem{7}  M. D'Agostino {\it et al}., Phys Lett. {\bf B473}, 219 (2000).
\bibitem{8}  L. Beaulieu {\it et al}., Phys. Rev. Lett. {\bf 84}, 5971 (2000).
\bibitem{9}  O. Lopez, Nucl. Phys. {\bf A685}, 246c (2001).
\bibitem{10}  B. Borderie {\it et al}., Phys. Rev. Lett. {\bf 86}, 3252 (2001).
\bibitem{11}  N.T. Porile {\it et al}., Phys. Rev. C {\bf 39}, 1914 (1989).
\bibitem{12}  S.P. Avdeyev {\it et al}., Yad. Fiz. {\bf 64}, 1628 (2001) (Phys. of Atom.
Nuclei {\bf 64}, 1549 (2001)). S.P. Avdeyev{\it et al}., Nucl. Phys. {\bf A709}, 392 (2002).
\bibitem{13}  A. Bonasera, M. Bruno, C.O. Dorso, and P.F. Mastinu, La Rivista
del Nuovo Chimento {\bf 23}, 2, 1 (2000).
\bibitem{14}  J. Richert and P. Wagner, Phys. Reports {\bf 350}, 1, 1 (2001).
\bibitem{15}   P. Bonche, S. Levit, and D. Vautherin, Nucl.Phys. {\bf A436}, 265 (1985).
\bibitem{16}  Feng-Shou Zhang, Z. Phys. {\bf A356}, 163 (1996).
\bibitem{17}  V.A. Karnaukhov {\it et al}.,Yad. Fiz. {\bf 62}, 272 (1999) (Phys. of At.
Nuclei {\bf 62}, 237 (1999)).
\bibitem{18}  A.S. Hirsch, A. Bujak, J.E. Finn, L.J. Gutay, N.T. Porile, R.P. Scharenberg, B.C. Stringfellow, and F. Turkot, Phys.Rev. C {\bf 29}, 508 (1984).
\bibitem{19}  M.E. Fisher, Physics {\bf 3}, 255 (1967).
\bibitem{20}  A.D. Panagiotou, M.W. Curtin, and D.K. Scott, Phys.Rev. C {\bf 31}, 55 (1985).
\bibitem{21}  J.B. Elliott, L.G. Moretto, L. Phair {\it et al}., Phys.Rev.Lett. {\bf 88}, 042701 (2002).
\bibitem{22}  J.B. Elliott {\it et al}., nucl-ex /0205004v1 (2002).
\bibitem{23}  V.A. Karnaukhov, Yad. Fiz. {\bf 60}, 1780 (1997) (Phys. of At.
Nuclei {\bf 60}, 1625 (1997).
\bibitem{24}  L.G. Moretto, S.G. Thompson, J. Routti, and R.C. Gatti, Phys. Letter {\bf B38}, 471 (1972).
\bibitem{25}  M. D'Agostino {\it et al}., Nucl. Phys. {\bf A650}, 329 (1999).
\bibitem{26}  M. Kleine Berkenbusch {\it et al}., Phys. Rev. Lett. {\bf 88}, 022701 (2002).
\bibitem{27}  A.S. Botvina, A.S. Il'inov, and I.N. Mishustin, Yad. Fiz. {\bf 42}, 1127 (1985) (Sov. J. Nucl. Phys. {\bf 42}, 712 (1985)).\\
J. Bondorf, R. Donangelo, I.N. Mishustin, and H. Schulz, Nucl. Phys. {\bf A444}, 460 (1985). 
\bibitem{28}  J. Bondorf, A.S. Botvina, A.S. Iljinov, I.N. Mishustin, and K. Sneppen, Phys. Rep. {\bf 257}, 133 (1995).
\bibitem{29}  R.P. Scharenberg {\it et al.}, Phys. Rev. C {\bf 64}, 054602 (2001).
\bibitem{30}  B.K. Srivastava {\it et al.}, Phys. Rev. C {\bf 65}, 054617 (2002).
\bibitem{31}  J. Schmelzer, G. R\"opke, and F.- P. Ludwig, Phys. Rev. C {\bf 55}, 1917 (1997).
\bibitem{32}  P.T. Reuter and K.A. Bugaev, Phys. Letter {\bf B517}, 233 (2001).
\bibitem{33}  V.D. Toneev, N.S. Amelin, K.K. Gudima, and S.Yu. Sivoklokov, Nucl. Phys. {\bf A519}, 463c (1990).
N.S. Amelin, K.K. Gudima, S.Yu. Sivoklokov, and V.D. Toneev, Yad. Fiz. {\bf 52}, 272 (1990).
\bibitem{34}  D.G. Ravenhall, C.J. Pethick, and J.M. Lattimer, Nucl. Phys.
{\bf A407}, 571 (1983).
\bibitem{35}  K.A. Bugaev, M.I. Gorenstein, I.N. Mishustin, and W. Greiner, Phys. Rev. C {\bf 62}, 044320 (2000).
\end{thebibliography}
\end{document}